# Optimal rate list decoding via derivative codes

Venkatesan Guruswami[*]     Carol Wang[†]

Computer Science Department
Carnegie Mellon University
Pittsburgh, PA 15213

**Abstract**

The classical family of $[n,k]_q$ Reed-Solomon codes over a field $\mathbb{F}_q$ consist of the evaluations of polynomials $f \in \mathbb{F}_q[X]$ of degree $< k$ at $n$ distinct field elements. In this work, we consider a closely related family of codes, called (order $m$) *derivative codes* and defined over fields of large characteristic, which consist of the evaluations of $f$ as well as its first $m-1$ formal derivatives at $n$ distinct field elements. For large enough $m$, we show that these codes can be list-decoded in polynomial time from an error fraction approaching $1-R$, where $R = k/(nm)$ is the rate of the code. This gives an alternate construction to folded Reed-Solomon codes for achieving the optimal trade-off between rate and list error-correction radius.

Our decoding algorithm is linear-algebraic, and involves solving a linear system to interpolate a multivariate polynomial, and then solving another structured linear system to retrieve the list of candidate polynomials $f$. The algorithm for derivative codes offers some advantages compared to a similar one for folded Reed-Solomon codes in terms of efficient unique decoding in the presence of side information.

## 1 Introduction

Consider the task of communicating information via transmission of $n$ symbols from a large alphabet $\Sigma$ over an adversarial channel that can arbitrarily corrupt any subset of up to $pn$ symbols (for some error parameter $p \in (0,1)$). Error-correcting codes can be used to communicate reliably over such a channel. A code $C$ is a judiciously chosen subset of $\Sigma^n$ that enables recovery of any $\mathbf{c} \in C$ from its distorted version $\mathbf{c} + \mathbf{r}$ so long as $\mathbf{r}$ has at most $pn$ nonzero entries. The rate $R$ of the code $C$ equals $\frac{\log |C|}{n \log \Sigma}$, which is the ratio of number of bits of information in the message to the total number of bits transmitted. A basic trade-off in this setting is the one between rate $R$ and error fraction $p$. Clearly, $R \leq 1 - p$, since the channel can always zero-out the last $pn$ symbols.

---
[*]Supported in part by a Packard Fellowship and NSF grants CCF 0953155 and CCF 0963975. Email: guruswami@cmu.edu

[†]Supported in part by NSF CCF 0963975 and MSR-CMU Center for Computational Thinking. wangc@cs.cmu.edu

## 1.1 Background

Perhaps surprisingly, the above simple limit can in fact be met, in the model of list decoding. Under list decoding, the error-correction algorithm is allowed to output a list of all codewords within the target error bound $pn$ from the noisy received word. If this output list-size is small, say a constant or some polynomially growing function of the block length, then this is still useful information to have in the worst-case instead of just settling for decoding failure. For a survey of algorithmic results in list decoding, see [4].

List decoding allows one to decode from an error fraction approaching the optimal limit of $1 - R$. In fact, there *exist* codes of rate $R$ that enable decoding up to a fraction $1 - R - \varepsilon$ of errors with a list-size bound of $O(1/\varepsilon)$ (this follows from standard random coding arguments). However, this is a nonconstructive result, with no deterministic way to construct a good code or an efficient algorithm to list decode it. Recently, it was shown that list decoding from an error rate approaching $1 - R$ is possible constructively, with an explicit code (the *folded Reed-Solomon code*) and a polynomial time decoding algorithm [8]. However, the list-size guarantee is much larger than the $O(1/\varepsilon)$ bound achieved by random codes, and is a large polynomial in the block length.

Before we state the result, let us first recall the definition of the well-known *Reed-Solomon codes*. For integer parameters $1 < k < n$, a field $\mathbb{F}$ of size $\geq n$, and a sequence $S = (a_1, \ldots, a_n)$ of $n$ distinct elements of $\mathbb{F}$, the associated Reed-Solomon (RS) code is

$$\mathrm{RS}_{\mathbb{F},S}[n, k] = \{(p(a_1), \ldots, p(a_n)) \mid p \in \mathbb{F}[X] \text{ of degree } < k\}.$$

The code $\mathrm{RS}_{\mathbb{F},S}[n, k]$ has rate $R = k/n$, and can be list-decoded from up to a $1 - \sqrt{R}$ fraction of errors [12, 9]. It is *not* known if list decoding some instantiation of Reed-Solomon codes from a larger radius is possible. At the same time, it is also not known if there are some RS codes for which the list-size could grow super-polynomially beyond this radius. For a more general problem called "list recovery," it is known that the error fraction cannot be improved for certain RS codes [7].

It turns out one can decode beyond the $1 - \sqrt{R}$ bound by augmenting the RS encoding with some extra information. Parvaresh and Vardy used the evaluations of polynomials carefully correlated with the message polynomial $p$ also in the encoding [11]. However, the encodings of the extra polynomial(s) cost a lot in terms of rate, so their improvement is confined to low rates (at most $1/16$) and does not achieve the optimal $1 - R$ radius. Later, Guruswami and Rudra considered a "folded" version of RS codes [8], which is really just the RS code viewed as a code over a larger alphabet. More precisely, the order-$m$ folded Reed-Solomon code is defined as follows.

**Definition 1.** Let $\mathbb{F}$ be a field of size $q$ with nonzero elements $\{1, \gamma, \ldots, \gamma^{n-1}\}$ for $n = q - 1$, where $\gamma$ is a primitive element of $\mathbb{F}$. Let $m \geq 1$ be an integer which divides $n$. Let $1 \leq k < n$ be the degree parameter.

The folded Reed-Solomon code $\mathrm{FRS}_{\mathbb{F}}^{(m)}[k]$ is a code over alphabet $\mathbb{F}^m$ that encodes a polynomial $f \in \mathbb{F}[X]$ of degree $k$ as

$$f(X) \mapsto \left( \begin{bmatrix} f(1) \\ f(\gamma) \\ \vdots \\ f(\gamma^{m-1}) \end{bmatrix}, \begin{bmatrix} f(\gamma^m) \\ f(\gamma^{m+1}) \\ \vdots \\ f(\gamma^{2m-1}) \end{bmatrix}, \ldots, \begin{bmatrix} f(\gamma^{n-m}) \\ f(\gamma^{n-m+1}) \\ \vdots \\ f(\gamma^{n-1}) \end{bmatrix} \right). \quad (1)$$

It is shown in [8] that the above code can be decoded up to an error fraction $\approx 1 - \left(\frac{mR}{m-s+1}\right)^{\frac{s}{s+1}}$ for any parameter $s$, $1 \leq s \leq m$, where $R = k/n$ is the rate of the code. (For $s = 1$, the performance ratio is the $1 - \sqrt{R}$ bound, but the radius improves for large $s$ and $m \gg s$. For example, picking $s \approx 1/\varepsilon$ and $m \approx 1/\varepsilon^2$, the list decoding radius exceeds $1 - R - \varepsilon$.) The bound on list-size is $q^{s-1}$, and the decoding complexity is of the same order. Getting around this exponential dependence on $s$ remains an important theoretical question.

The above algorithm involved finding roots of a univariate polynomial over an extension field of large degree over the base field $\mathbb{F}$. Recently, an entirely linear-algebraic algorithm was discovered in [6] which avoids the use of extension fields. Although the error fraction decoded by the linear-algebraic algorithm is smaller — it is $\frac{s}{s+1}\left(1 - \frac{mR}{m-s+1}\right)$ for the above folded RS codes — it can still be made to exceed $1 - R - \varepsilon$ for any $\varepsilon > 0$ by the choices $s \approx 1/\varepsilon$ and $m \approx 1/\varepsilon^2$. The advantage of the algorithm in [6] is that except for the step of pruning an $(s-1)$-dimensional subspace to filter the close-by codewords, it has quadratic running time.

## 1.2 This work

In this work, we consider another natural variant of Reed-Solomon codes (over fields of large characteristic), called *derivative codes*, defined formally in Section 2. Informally, rather than bundling together evaluations of the message polynomial at consecutive powers of $\gamma$, in an order-$m$ derivative code, we bundle together the evaluations of $f$ as well as its first $(m-1)$ derivatives at each point. This might appear to cause a loss in rate (similar to the Parvaresh-Vardy construction [11]), but it does not, as one can pick higher degree polynomials while still maintaining the distance. (For two distinct degree $\ell$ polynomials, there can be at most $\ell/m$ points where they and their first $(m-1)$ derivatives agree.)

In Theorem 6 and Corollary 7, we show our main result that derivative codes also achieve list-decoding capacity; that is, for any $\varepsilon > 0$, for the choice $m \approx 1/\varepsilon^2$, we can list decode order-$m$ derivative codes of rate $R$ from a $1 - R - \varepsilon$ fraction of errors. The list-size and running time behavior is similar to the linear-algebraic algorithm for folded RS codes [6], and once again one can find, by just solving two linear systems, a low-dimensional space that contains all the close-by codewords.

Recently, multivariate versions of derivative codes were used in [10] to give locally decodable codes. In that work, these codes were referred to as *multiplicity codes*, but we refer to our codes as *derivative codes* to emphasize our use of formal derivatives rather than Hasse derivatives in the encoding. A side benefit of the changed terminology is to single out the important univariate case with a different name.

**Motivation.** Prior to this work, the only known explicit codes list decodable up to the optimal $1 - R$ bound were based on folded Reed-Solomon codes (or with smaller alphabets, certain folded algebraic-geometric codes [5], though these are not fully explicit). It seems like a natural question to seek alternate algebraic constructions of such codes. In addition, there is the possibility that a different construction would have better complexity or list-size guarantees, or offer other advantages.

The derivative code construction is arguably just as natural as the folded Reed-Solomon one.

Interestingly, it falls in the framework of Parvaresh-Vardy codes, where the correlated polynomials are formal derivatives. The special properties of derivatives ensures that one need not suffer any loss in rate, and at the same time enable list decoding up to a much larger radius than the bound for RS codes. Further, our algorithm for list decoding derivative codes has some nice properties with respect to decoding with side information, and might have some benefits in practice as well. However, as with the case of folded RS codes, the proven bound on the worst-case list size has an exponential dependence on $\varepsilon$ (when the decoding radius is $1 - R - \varepsilon$), and it remains a challenge to improve this. We should note that we cannot rule out the possibility that a better analysis can improve the bound; in general it is a very hard problem to show list-size *lower bounds* for these algebraic codes.

We end the introduction with a brief overview of the algorithm, and speculate on a possible benefit it offers compared to the folded RS case. At a high level, our decoding algorithm is similar to those used for Reed-Solomon and folded Reed-Solomon codes — it consists of an interpolation step, and then a second step to retrieve the list of all polynomials satisfying a certain algebraic condition. The interpolation step consists of fitting a polynomial of the form $A_0(X) + A_1(X)Y_1 + A_2(X)Y_2 + \cdots + A_s(X)Y_s$. (Note that the total degree in the $Y_i$'s is 1, and we do not use "multiplicities" in the interpolation.) The second step consists of solving the "differential equation" $A_0(X) + A_1(X)f(X) + A_2(X)f'(X) + \ldots + A_s(X)f^{(s-1)}(X) = 0$ for low-degree polynomials $f$. (Independently, a list decoding guarantee similar to the Guruswami-Rudra bound for folded RS codes has been obtained by Bombieri and Kopparty [1] based on using higher powers of $Y_i$ as well as multiplicities in the interpolation.)

The differential equation imposes a system of linear equations on the coefficients of $f$. The specific structure of this linear system is different from the one for folded RS codes in [6]. In particular, once the values of $f$ and its first $s-2$ derivatives at some point $\alpha$ (at which the interpolated polynomial $A_s$ doesn't vanish) are known, the rest are determined by the system. This has two advantages. First, having these values (at a random $\alpha$) as side information immediately leads to an efficient unique decoding algorithm. Second, in practice, $A_s$ may not have many zeroes amongst the evaluation points, in which case we can obtain the values of $f(a_i), \ldots, f^{(s-2)}(a_i)$ from the received word (instead of trying all $q^{s-1}$ possibilities). While we have not been able to leverage this structure to improve the worst-case list-size bound, it is conceivable that additional ideas could lead to some improvements.

## 2 Derivative codes

We denote by $\mathbb{F}_q$ the field of $q$ elements. For a polynomial $f \in \mathbb{F}_q[X]$, we denote by $f'$ its formal derivative, i.e. if $f(X) = f_0 + f_1 X + \ldots + f_\ell X^\ell$, then $f'(X) = \sum_{i=1}^{\ell} i f_i X^{i-1}$. We denote by $f^{(i)}$ the formal $i$'th derivative of $f$.

**Definition 2** (*m*'th order derivative code). Let $0 \leq m \in \mathbb{Z}$. Let $a_1, \ldots, a_n \in \mathbb{F}_q$ be distinct, and let the parameters satisfy $m \leq k < nm \leq q$. Further assume that $\text{char}(\mathbb{F}_q) > k$.

The derivative code $\text{Der}_q^{(m)}[n,k]$ over the alphabet $\mathbb{F}_q^m$ encodes a polynomial $f \in \mathbb{F}_q[X]$ of

degree $k-1$ by

$$f \mapsto \left( \begin{bmatrix} f(a_1) \\ f'(a_1) \\ \vdots \\ f^{(m-1)}(a_1) \end{bmatrix}, \begin{bmatrix} f(a_2) \\ f'(a_2) \\ \vdots \\ f^{(m-1)}(a_2) \end{bmatrix}, \ldots, \begin{bmatrix} f(a_n) \\ f'(a_n) \\ \vdots \\ f^{(m-1)}(a_n) \end{bmatrix} \right). \tag{2}$$

*Remark* 1. Note that the case $m = 1$ is a Reed-Solomon code.

This code has block length $n$ and rate $R = \frac{k}{nm}$. The minimum distance is $n - \lfloor \frac{k-1}{m} \rfloor \approx (1-R)n$.

## 3 List decoding derivative codes

Suppose we have received the corrupted version of a codeword from the derivative code $\mathrm{Der}_q^{(m)}[n,k]$ as a string $\mathbf{y} \in (\mathbb{F}_q^m)^n$, which we will naturally consider as an $m \times n$ matrix over $\mathbb{F}_q$:

$$\begin{pmatrix} y_{11} & y_{12} & \cdots & y_{1n} \\ y_{21} & y_{22} & \cdots & y_{2n} \\ \vdots & \vdots & \ddots & \vdots \\ y_{m1} & y_{m2} & \cdots & y_{mn} \end{pmatrix}. \tag{3}$$

The goal is to recover all polynomials $f$ of degree $k-1$ whose encoding (2) agrees with $\mathbf{y}$ in at least $t$ columns. This corresponds to decoding from $n-t$ symbol errors for the derivative code $\mathrm{Der}_q^{(m)}[n,k]$. When $t > (n+k/m)/2$, the polynomial $f$, if it exists, is unique, and in this regime an efficient decoding algorithm was given in [10] by adapting the Welch-Berlekamp algorithm for Reed-Solomon codes [14, 2].

We adapt the algebraic list-decoding method used for Reed-Solomon and folded Reed-Solomon codes to the derivative code setting. The decoding algorithm consists of two steps — (i) interpolation of an algebraic condition (that must be obeyed by all candidate polynomials $f$), and (ii) retrieving the list of candidate solutions $f$ (from the algebraic condition found by the interpolation step).

Our algorithm can be viewed as a higher dimensional analog of the Welch-Berlekamp algorithm, where we use multivariate polynomials instead of bivariate polynomials in the interpolation. This has been used in the context of folded Reed-Solomon codes in [13, Chap. 5] and [6], and here we show that derivative codes can also be list decoded in this framework.

### 3.1 Interpolation

Let $\mathcal{W}$ denote the $\mathbb{F}_q$-linear subspace of $\mathbb{F}_q[X, Y_1, \ldots, Y_m]$ consisting of polynomials that have total degree at most 1 in the $Y_i$'s, i.e, $\mathcal{W}$ contains polynomials of the form $B_0(X) + B_1(X)Y_1 + B_2(X)Y_2 + \cdots + B_m(X)Y_m$ for some polynomials $B_i \in \mathbb{F}_q[X]$.

Let $D$ be the $\mathbb{F}_q$-linear map on $\mathcal{W}$ defined as follows: For $p \in \mathbb{F}_q[X]$, and $1 \le i \le m$,

$$D(p)(X, Y_1, \ldots, Y_m) = p'(X) \tag{4}$$

and
$$D(pY_i)(X, Y_1, \ldots, Y_m) = p'(X)Y_i + p(X)Y_{i+1}. \tag{5}$$

where we take $Y_{m+1} = Y_1$.

Let $s$, $1 \leq s \leq m$, be an integer parameter in the decoding algorithm. The goal in the interpolation step is to interpolate a **nonzero** polynomial $Q \in \mathbb{F}_q[X, Y_1, Y_2, \ldots, Y_s]$ of the form

$$A_0(X) + A_1(X)Y_1 + A_2(X)Y_2 + \cdots + A_s(X)Y_s \tag{6}$$

satisfying the following conditions for each $i$, $1 \leq i \leq n$:

$$Q(a_i, y_{1i}, \ldots, y_{si}) = 0 \text{ and } (D^k Q)(a_i, y_{1i}, \ldots, y_{mi}) = 0 \quad (k = 1, \ldots, m-s), \tag{7}$$

where $D^k$ denotes the $k$-fold composition of the map $D$.

**Observation.** *For each $i$, the conditions (7) are a collection of $(m-s+1)$ homogeneous linear constraints on the coefficients of the polynomial $Q$.*

The following shows why the interpolation conditions are useful in the decoding context.

**Lemma 1.** *Suppose $Q$ of the form (6) satisfies the conditions (7). If the received word (3) agrees with the encoding of $f$ at location $i$, that is, $f^{(j)}(a_i) = y_{j+1,i}$ for $0 \leq j < m$, then the univariate polynomial $\hat{Q}(X) := Q(X, f(X), \ldots, f^{(s-1)}(X))$ satisfies $\hat{Q}(a_i) = 0$ as well as $\hat{Q}^{(k)}(a_i) = 0$ for $k = 1, \ldots, m-s$, where $\hat{Q}^{(k)}(X)$ is that the k'th derivative of $\hat{Q}$.*

*Proof.* Notice the form that our definition of the map $D$ takes when $Y_i = f^{(i-1)}(X)$ for $1 \leq i \leq m$. We have $D(p) = p'$ for $p \in \mathbb{F}_q[X]$, and $D(pf^{(i-1)}) = p'f^{(i-1)} + pf^{(i)}$, which is simply the product rule for derivatives. Thus when $(y_{1i}, y_{2i}, \ldots, y_{mi}) = (f(a_i), f'(a_i), \ldots, f^{(m-1)}(a_i))$, the conditions (7) enforce that $\hat{Q}$ and its first $m-s$ derivatives vanish at $a_i$. □

We next argue that a nonzero interpolation polynomial $Q$ exists and can be found efficiently.

**Lemma 2.** *Let*
$$d = \left\lfloor \frac{n(m-s+1) - k + 1}{s+1} \right\rfloor. \tag{8}$$

*Then, a nonzero $Q$ of the form (6) satisfying the conditions (7) with $\deg(A_0) \leq d + k - 1$ and $\deg(A_j) \leq d$ for $1 \leq j \leq s$ exists and can be found in $O((nm)^3)$ field operations over $\mathbb{F}_q$.*

*Proof.* Under the stated degree restrictions, the number of monomials in $Q$ is

$$(d+1)s + d + k = (d+1)(s+1) + k - 1 > n(m-s+1).$$

where the last inequality follows from the choice (8) of $d$. The number of homogeneous linear equations imposed on the coefficients of $Q$ in order to meet the interpolation conditions (7) is $n(m-s+1)$. As this is less than the number of monomials in $Q$, the existence of a nonzero $Q$ follows, and it can be found by solving a linear system over $\mathbb{F}_q$ with at most $nm$ constraints. □

## 3.2 Retrieving candidate polynomials

Suppose we have a polynomial $Q(X, Y_1, \ldots, Y_s)$ satisfying the interpolation conditions (7). The following lemma gives an identity satisfied by any $f$ which has good agreement with the received word.

**Lemma 3.** *If $f \in \mathbb{F}[X]$ has degree at most $k - 1$ and an encoding (2) agreeing with the received word $\mathbf{y}$ in at least $t$ columns for $t > \frac{d+k-1}{m-s+1}$, then*

$$Q(X, f(X), f'(X), \ldots, f^{(s-1)}(X)) = 0.$$

*Proof.* Let $\hat{Q}(X) = Q(X, f(X), \ldots, f^{(s-1)}(X))$. By Lemma 1, an agreement in column $i$ means that $\hat{Q}(X)$ satisfies $\hat{Q}(a_i) = 0$ and that the $k$th derivative $\hat{Q}^{(k)}(a_i)$ is also zero for $k = 1, \ldots, m - s$. In particular, $t$ column agreements yield at least $t(m - s + 1)$ roots (counting multiplicities) for $\hat{Q}$.

The degree of $\hat{Q}$ is at most $d + k - 1$, as $f$ and each of its derivatives has degree at most $k - 1$. Then as $\hat{Q}$ is univariate of degree at most $d + k - 1$, $\hat{Q}$ has at most $d + k - 1$ roots if it is nonzero. Thus if $t > (d + k - 1)/(m - s + 1)$, it must be that $\hat{Q}(X) = 0$. □

With our chosen value of $d$ from (8), this means that any $f$ which agrees with $\mathbf{y}$ on more than

$$\frac{n}{s+1} + \frac{s}{s+1}\frac{k-1}{m-s+1} \tag{9}$$

columns satisfies $Q(X, f(X), f'(X), \ldots, f^{(s-1)}(X)) = 0$. So in the second step, our goal is to find all polynomials $f$ of degree at most $k - 1$ such that

$$A_0(X) + A_1(X)f(X) + A_2(X)f'(X) + \ldots + A_s(X)f^{(s-1)}(X) = 0 \tag{10}$$

Let $A_i(X) = \sum_{j=0}^{\deg(A_i)} a_{ij} X^j$ for each $i$. Note that the above constraint (10) gives a *linear system* over $\mathbb{F}$ in the coefficients of $f = f_0 + f_1 X + \cdots + f_{k-1} X^{k-1}$. In particular, the set of solutions $(f_0, f_1, \ldots, f_{k-1})$ is an affine space, and we can find it by solving the linear system. Our goal now is to bound the dimension of the space of solutions by exposing its special structure and also use this to efficiently find an explicit basis for the space.

**Lemma 4.** *It suffices to give an algorithm in the case that the constant term $a_{s0}$ of $A_s$ is nonzero.*

*Proof.* If $A_s(X) \not\equiv 0$, since $\deg(A_s) \le d < nm \le q$, then there is some $\alpha \in \mathbb{F}_q$ such that $A_s(\alpha) \ne 0$, so we can consider a "translate" of this problem by $\alpha$; that is, $A_s(X + \alpha)$ has nonzero constant term, so we can solve the system with the translated polynomial $Q(X + \alpha, Y_1, \ldots, Y_m)$ and recover candidate messages by translating each solution $g(X)$ to $f(X) = g(X - \alpha)$.

If $A_s(X) = 0$, we simply reduce the problem to a smaller one with $s$ rather than $s + 1$ interpolation variables. Note that this must terminate since $Q$ is nonzero and so at least one $A_i$ for $i \ge 1$ is nonzero. □

We can now show:

**Lemma 5.** *If $a_{s0} \ne 0$, the solution space to (10) has dimension at most $s - 1$.*

*Proof.* For each power $X^i$, the coefficient of $X^i$ in $A_0(X) + A_1(X)f(X) + \cdots + A_s(X)f^{(s-1)}(X)$ is

$$a_{0i} + \big(a_{10}f_i + a_{11}f_{i-1} + \cdots + a_{1i}f_0\big) + \big(a_{20}(i+1)f_{i+1} + a_{21}if_i + \cdots + a_{2i}f_1\big)$$
$$+ \cdots + \big(a_{s0}(i+s-1)(i+s-2)\cdots(i+1)f_{i+s-1} + \cdots + a_{si}(s-1)!f_{s-1}\big)$$
$$= a_{0i} + \sum_{j=1}^{s}\sum_{k=0}^{i} \frac{(k+j-1)!}{k!} a_{j(i-k)} f_{k+j-1}.$$

If $(f_0, \ldots, f_{k-1})$ is a solution to (10), then this coefficient is zero for every $i$.

The coefficient of $X^i$ for each $i$ depends only on $f_j$ for $j < i+s$, and the coefficient of $f_{i+s-1}$ is $a_{s0}(i+s-1)(i+s-2)\cdots(i+1)$, which is nonzero when $i+s \leq k$ since $\mathrm{char}(\mathbb{F}_q) > k$. Thus, if we fix $f_0, f_1, \ldots, f_{s-2}$, the rest of the coefficients $f_{s-1}, \ldots, f_{k-1}$ are uniquely determined. In particular, the dimension of the solution space is at most $s-1$. □

*Remark* 2. The bound of Lemma 5 is tight for arbitrary linear systems. Indeed, if

$$Q(X, Y_1, \ldots, Y_s) = \sum_{i=0}^{s-1} \frac{(-1)^i}{i!} X^i Y_{i+1},$$

then *any* polynomial of degree less than $s$ with zero constant term satisfies $Q(X, f(X), \ldots, f^{(s-1)}(X)) = 0$. This is because any monomial $f(X) = X^j$ for $0 < j \leq s-1$ is a solution, and our solution space is linear. Of course, we do not know if such a bad polynomial can occur as the output of the interpolation step when decoding a noisy codeword of the derivative code.

Combining these lemmas and recalling the bound (9) on the number of agreements for successful decoding, we have our main result.

**Theorem 6** (Main). *For every $1 \leq s \leq m$, the derivative code $\mathrm{Der}_q^{(m)}[n,k]$ (where $\mathrm{char}(\mathbb{F}_q) > k$) satisfies the property that for every received word $\mathbf{y} \in \mathbb{F}_q^{nm}$, an affine subspace $S \subseteq \mathbb{F}_q[X]$ of dimension at most $s-1$ can be found in polynomial time such that every $f \in \mathbb{F}_q[X]$ of degree less than $k$ whose derivative encoding differs from $\mathbf{y}$ in at most*

$$\frac{s}{s+1}\left(n - \frac{k}{(m-s+1)}\right)$$

*positions belongs to $S$.*

Now by setting $s \approx 1/\varepsilon$ and $m \approx 1/\varepsilon^2$, and recalling that the rate of $\mathrm{Der}_q^{(m)}[n,k]$ equals $k/(nm)$, we can conclude the following.

**Corollary 7.** *For all $R \in (0,1)$ and all $\varepsilon > 0$, for a suitable choice of parameters, there are derivative codes $\mathrm{Der}_q^{(m)}[n,k]$ of rate at least $R$ which can be list decoded from a fraction $1 - R - \varepsilon$ of errors with a list-size of $q^{O(1/\varepsilon)}$.*

## 4 Some remarks

We now make a couple of remarks on coping with the large list-size bound in our decoding algorithms.

## 4.1 Reducing the list size

One approach to avoid the large list size bound of $\approx q^s$ for the number of codewords near $f$ is to draw codewords from so-called *subspace-evasive* subsets of $\mathbb{F}_q^k$ rather than all of $\mathbb{F}_q^k$. This approach was used in [6] to reduce the list-size for folded Reed-Solomon codes, and we can gain a similar benefit in the context of list decoding derivative codes. A subset of $\mathbb{F}_q^k$ is $(s, L)$-subspace-evasive if it intersects with every linear subspace $S \subseteq \mathbb{F}_q^k$ of dimension at most $s$ in at most $L$ points.

For any $\varepsilon > 0$, a probabilistic argument shows that there exist $(s, O(s/\varepsilon))$-subspace-evasive subsets of $\mathbb{F}_q^k$ of size $q^{(1-\varepsilon)k}$. In fact, we have the following stronger statement, proved in [6]. Fix a basis $1, \beta, \ldots, \beta^{k-1}$ of $\mathbb{F}_q^k$ over $\mathbb{F}_q$ and denote $\mathbb{K} = \mathbb{F}_{q^k}$. For $P \in \mathbb{K}[X]$ and an integer $r$, $1 \le r \le k$, define

$$\mathbb{S}(P, r) \;=\; \{(a_0, \ldots, a_{k-1}) \in \mathbb{F}_q^k \mid P(a_0 + a_1\beta + \cdots + a_{k-1}\beta^{k-1}) \in \mathbb{F}_q\text{-span}(1, \beta, \ldots, \beta^{r-1})\}.$$

**Lemma 8** ([6]). *Let $q$ be a prime power, $k \ge 1$ an integer. Let $\zeta \in (0, 1)$ and $s \in \mathbb{Z}$ satisfying $1 \le s \le \zeta k/2$. Let $P \in \mathbb{K}[X]$ be a random polynomial of degree $t$ and define $\mathcal{V} = \mathbb{S}(P, (1-\zeta)k)$. Then for $t \ge \Omega(s/\zeta)$, with probability at least $1 - q^{-\Omega(k)}$ over the choice of $P$, $\mathcal{V}$ is an $(s, t)$-subspace-evasive subset of $\mathbb{F}_q^k$ of size at least $q^{(1-\zeta)k}/2$.*

By taking messages from $\mathcal{V}$ rather than all of $\mathbb{F}_q^k$, we suffer a small loss in rate, but give a substantial improvement to the list size bound; since our solution space is linear, the number of candidate messages is reduced from $\approx q^s$ to $O(s/\varepsilon)$. In particular, setting our parameters as in Theorem 6, we can list-decode from a $1 - R - \varepsilon$ fraction of errors with a list size of at most $O(1/\varepsilon^2)$. However, the code construction is not explicit but only a randomized (Monte Carlo) one that satisfies the claimed guarantees on list-decoding with high probability.

## 4.2 Decoding with side information

The decoding described in the previous section consists of trying all choices for the coefficients $f_0, \ldots, f_{s-2}$ and using each to uniquely determine a candidate for $f$. Note however that for each $i$, the $f_i$ is essentially the $i$th derivative of $f$ evaluated at $0$, and can be recovered as $f^{(i)}(0)/i!$. Thus if the decoder somehow knew the correct values of $f$ and its first $s - 1$ derivatives at $0$, $f$ could be recovered uniquely (as long as $A_s(0) \ne 0$).

Now, suppose the encoder could send a small amount of information along a noiseless side channel in addition to sending the (much longer) codeword on the original channel. In such a case, the encoder could choose $\alpha \in \mathbb{F}_q$ uniformly at random and transmit $f(\alpha), f'(\alpha), \ldots, f^{(s-1)}(\alpha)$ on the noiseless channel. The decoding then fails only if $A_i(\alpha) = 0$ for $i$ which is the largest index such that $A_i(X) \ne 0$. As the $A_i(X)$ have bounded degree, by increasing the field size $q$, $f$ can be uniquely recovered with probability arbitrarily close to 1. More precisely, we have the following claim.

**Theorem 9.** *Given a uniformly random $\alpha \in \mathbb{F}_q$ and the values $f(\alpha), f'(\alpha), \ldots, f^{(s-1)}(\alpha)$ of the message polynomial $f$, the derivative code $\mathrm{Der}_q^{(m)}[n, k]$ can be uniquely decoded from up to*

$$\frac{s}{s+1}\left(n - \frac{k}{m-s+1}\right)$$

*errors with probability at least $1 - \frac{nm}{sq}$ over the choice of $\alpha$.*

*Proof.* As in the proof of Lemma 4, as long as $A_s(\alpha) \neq 0$, we may translate the problem by $\alpha$ and use the values $f(\alpha), f'(\alpha), \ldots, f^{(s-1)}(\alpha)$ to uniquely determine the shifted coefficients $g_0, \ldots, g_{s-1}$.

As $A_s \neq 0$, and $A_s$ is univariate of degree at most $d$, $A_s$ has at most $d$ roots, and so the probability that $A_s(\alpha) \neq 0$ is at least $1 - d/q \geq 1 - \frac{nm}{sq}$, where the last inequality follows from our choice of $d \leq nm/s$ in (8). □

*Remark* 3. In the context of communicating with side information, there is a generic, black-box solution combining list-decodable codes with hashing to guarantee unique recovery of the correct message with high probability [3]. In such a scheme, the side information consists of a random hash function $h$ and its value $h(f)$ on the message $f$. The advantage of the solution in Theorem 9 is that there is *no need* to compute the full list (which is the computationally expensive step, since the list size bound depends exponentially on $s$) and then prune it to the unique solution. Rather, we can uniquely identify the first $(s-1)$ coefficients of the polynomial $f(X + \alpha)$ in the linear system (10), after applying the shift $X \mapsto X + \alpha$, as $f(\alpha), f'(\alpha), \ldots, f^{(s-2)}(\alpha)$. Then, as argued in the proof of Lemma 5, the remaining coefficients are determined as linear combinations of these $s - 1$ coefficients. So the whole algorithm can be implemented in quadratic time.

*Remark* 4. The decoder could use the columns of the received word **y** as a guess for the side information $f(a_i), f'(a_i), \ldots, f^{(s-2)}(a_i)$ for $i = 1, 2, \ldots, n$. Since $f$ agrees with **y** on more than $t > Rn$ positions, as long as $A_s(a_i) = 0$ for less than $t$ of the evaluation points $a_i$, we will recover every solution $f$ this way. This would lead to a list size bound of at most $n - t < n$. Unfortunately, however, there seems to be no way to ensure that $A_s$ does not vanish at most (or even all) of the points $a_i$ used for encoding. But perhaps some additional ideas can be used to make the list size polynomial in both $q, s$, or at least $\exp(O(s))q^c$ for some absolute constant $c$.